\documentclass[reprint, superscriptaddress,amsmath,amssymb,aps]{revtex4-2}
\usepackage[utf8]{inputenc}

\usepackage{graphicx}
\usepackage{physics}
\usepackage[english]{babel}
\usepackage{dcolumn}
\usepackage{bm}
\usepackage[dvipsnames]{xcolor}
\usepackage{hyperref}
\usepackage[mathlines]{lineno}

\begin{document}

\preprint{APS/123-QED}

\title{Propagation of optical vector vortices of slow light in a coherently prepared tripod configuration}

\author{Dharma P. Permana}
\email{dharma.permana@ff.stud.vu.lt}
\email{dhp.permana@gmail.com}
\affiliation{%
 Institute of Theoretical Physics and Astronomy, Vilnius University, Saul\.{e}tekio 3, Vilnius LT-10257, Lithuania
}%
\affiliation{Laser Research Center, Vilnius University, Vilnius, LT-10223, Lithuania}
\affiliation{D\'{e}partement Physique, Facult\'{e} des Sciences, Aix-Marseille Universit\'{e}, Marseille, France}

\author{Mazena Mackoit Sinkevi\v{c}ien\.{e}}%
 \email{mazena.mackoit-sinkeviciene@ff.vu.lt}
\affiliation{%
 Institute of Theoretical Physics and Astronomy, Vilnius University, Saul\.{e}tekio 3, Vilnius LT-10257, Lithuania
}%

\author{Julius Ruseckas}
\email{julius.ruseckas@gmail.com}
\affiliation{Baltic Institute of Advanced Technology, LT-01403 Vilnius, Lithuania}
\author{Hamid R. Hamedi}
 \email{hamid.hamedi@tfai.vu.lt}
\affiliation{%
 Institute of Theoretical Physics and Astronomy, Vilnius University, Saul\.{e}tekio 3, Vilnius LT-10257, Lithuania
}%



\date{\today}

\begin{abstract}
We investigate the propagation of optical vector vortices of slow light in a coherently prepared four-level tripod atomic system. The vector vortex consists of superposed pulse pairs with opposite circular polarizations and orbital angular momentum (OAM) charges $\pm l$, weakly interacting with an atomic medium initially prepared in a coherent superposition of two ground states. A third unoccupied state is coupled to a stronger control laser without OAM, creating a phase-dependent configuration. In the linear regime, the vortex OAM is mapped onto the medium, producing symmetrical azimuthally structured absorption patterns, with losses significantly reduced by the control field. For small detunings, complementary spatially dependent amplification and absorption occur for the two circular polarization components. This OAM-structured coherence induces a dynamical anisotropy, affecting both the intensity and polarization of the slow-light vortex. Polarization states evolve periodically between left-circular, linear, and right-circular polarizations during propagation. Once the beam reaches a stationary regime, the ring-shaped intensity transforms into a petal-like structure, and the final polarization states stabilize according to the initial superposition. The rate of polarization transitions is tunable via the control field strength, demonstrating flexible control over slow-light vector vortex dynamics.
  
\end{abstract}

\maketitle


\section{\label{Introduction}Introduction}
Over the past decades, intensive investigation of coherent pulse propagating through an ensemble of atomic media has led to the discovery of a rich variety of novel phenomena. One of the most important effects is electromagnetically induced transparency (EIT) \cite{boller1991,fleischhauer2000,Marangos.RMP2005}, which transforms an initially opaque medium into transparent in the presence of a strong control field that introduces quantum coherence between atomic states and leads to a modification of medium susceptibility \cite{paspalakis2002,Lukin.Nature2001,Finkelstein2023}. Since its discovery, a wide range of investigations have been carried out in an EIT-enabled atomic system, leading to the discovery of slow light propagation \cite{Hau1999,paspalakis2002,juzeliunas2004} where the speed of weak pulse propagation can be reduced to the order of tens of meters per second in the presence of a stronger control field. This demonstration of slow light in an EIT system paved the way for even wider discoveries of intriguing phenomena such as enhancement of optical nonlinearities \cite{minxiao2001,gong2006,hamedi2015}, stored light \cite{fleischhauer2000,lukin2001}, and stationary light \cite{Otterbach2010,Peters2022,Kim2022} with promising applications in the storage of quantum information carried by lights inside an atomic medium.

Studies of light-matter interaction within EIT-enabled atomic media can be further expanded to a structured light regime involving a special type of light carrying orbital angular momentum (OAM) known as the optical vortex \cite{coullet1989,brambilla1991,allen1992}. This special type of light beam is characterized by the possession of a helical phase term $e^{il\phi}$, where the topological charge $l$ is related to its discrete values of OAM $l\hbar$, which can take any integer value of $l$ \cite{brambilla1991,allen1992}. The boundless values of the topological charge open up a promising application for quantum information encoding in a higher dimension \cite{gibson2004,park2018,erhard2018,plachta2022}. Moreover, the combination of the unique properties of the vortex and its interaction with EIT-enabled atomic media has led to the discovery of intriguing phenomena such as azimuthally modulated transparency \cite{hamid2018} and the transfer of OAM in coherently prepared atomic media without control field \cite{Hamid.PRA2019}, and in the slow light regime with control field present \cite{ruseckas2013,Hamid2023}, which offers a way to store the OAM properties of vortices within the atomic media.

The potential of higher dimensional encoding offered by the OAM state of vortices can be enhanced further by a full vectorial consideration of optical vortices with spatially varying polarization states \cite{rosales2018}. This will allow one to access the spin angular momentum (SAM) states of a photon, encoded in its polarization texture \cite{zhang2010}, thus increasing the dimensionality even further. Furthermore, the interaction of optical vector vortices with EIT-enabled atomic media has led to the discovery of exotic phenomena such as spatially dependent transparency \cite{radwell2017,tarak2017}, and the transfer of SAMs of photons that manifested as an evolving spatial polarization structure \cite{Li.OE2022,Kudriasov.OE2025,Permana2025}. However, the slow light propagation regime of vector vortices has yet to be explored in these studies. Another crucial issue is that although spatially dependent transparency arises from the weak interaction of vector vortices with the EIT medium, the beam still experiences strong absorption in the azimuthally non-transparent regions upon entering the medium. 

In this work, we demonstrate the spatially dependent transparency and SAM exchange of vector vortices in the slow light propagation regime. We studied an optical vector vortex constructed from a pair of vortex pulses carrying opposite vorticity and orthogonal circular polarizations propagating in an ensemble of atomic media that was prepared in a superposition of two ground-states (phaseonium)\cite{scully.book1997}. The interaction of this pair of vortex pulses couple the phaseonium to an excited state and formed a coherently prepared $\Lambda$ configuration \cite{Permana2025}. Then a stronger control field is used to couple a third unoccupied ground state with the excited state of the $\Lambda$ system which transforms the configuration into a four-level tripod configuration. We derive the steady-state solution for the atomic coherences of this tripod configuration by assuming that the vortex pairs intensity is weaker than the control field. The analytical solutions of the atomic coherences, combined with the Maxwell–Bloch equations solved in the linear regime under the neglect of diffraction, reveal the mapping of the vortex OAM onto the spatial distribution of atomic coherence. This mapping leads to azimuthally dependent absorption with a $2|l|$-fold degeneracy, reduced absorption, and a positive dispersion slope, indicating a slow light propagation regime at resonance and complementary pairs of gain and loss with the same $2|l|$-fold degeneracy when the beam is small detuned from the resonance while still retaining a good performance of slow light propagation. The spatially structured coherence also introduces an effective anisotropy into the medium, which is manifested as an evolving polarization texture during beam propagation, where the final polarization state is determined from the choice of phaseonium component population regardless of the beam's initial polarization. Furthermore, increasing the strength of the control field allows one to slow 
the dynamics of polarization evolution and further reduce the absorption. Overall, this configuration established a novel interface for controlling the OAM and SAM dynamics of slow light vector vortices through atomic coherence and optical control.

\section{\label{Theoretical Model}Theoretical Formulation}
\subsection{\label{Atom-light configuration}Light–Matter Coupling Scheme}
We analyze a structured weak vector vortex probe field constructed from the coherent superposition of two pulse pairs carrying opposite orbital angular momentum (OAM) indices. The two modes correspond to right- and left-circularly polarized components. At the entrance plane of the atomic ensemble ($z=0$), the total electric field is written as
\begin{equation}
    \label{eq.1}
    \vec{E}(r,\phi,z=0) = E_R(r,\phi)\vec{e}_R+E_L(r,\phi)\vec{e}_L, 
\end{equation}
where $E_{R(L)}$ denotes the electric field amplitude for the right (left) -handed component of the vortex pairs as a function of radial position $r$ and azimuthal angle $\phi$ in the transverse plane of the beam. For analytical tractability, both components are assumed to follow Laguerre–Gaussian (LG) spatial modes of lowest radial index \cite{Kudriasov.OE2025}. The corresponding Rabi frequencies can then be written as
\begin{subequations}
\label{eq.2} 
 \begin{align}
     \Omega_R(r,\phi,z=0) &=\epsilon_R A(r)e^{il\phi}, \label{eq.2a} \\
    \Omega_L(r,\phi,z=0) &=\epsilon_L A(r) e^{-il\phi}, \label{eq.2b}
 \end{align}
\end{subequations}%
where $\Omega_R$ and $\Omega_L$ are the corresponding Rabi frequencies of the right- and left-circular components, and are related to the electric field via $\Omega_{R(L)}=\vec{d}_{R(L)}\cdot \vec{E}_{R(L)}/\hbar$, with $\vec{d}_{R(L)}$ representing the dipole moment of the atomic transition corresponding to the electric field component $\vec{E}_{R(L)}$ \cite{scully.book1997}. The function $A(r)$ represents the radial amplitude profile of the beam. For the lowest-radial-index Laguerre–Gaussian mode, it takes the form
\begin{equation}
    A(r) = \left(\frac{r}{w}\right)^{|l|}e^{-\frac{r^2}{w^2}}, \label{eq.3}
\end{equation}
where $w$ is the beam waist parameter. The weighting coefficients $\epsilon_R$ and $\epsilon_L$ determine the relative amplitudes of the two circular components. They are parameterized as
\begin{subequations}
\label{eq.4}
\begin{align}
    \epsilon_L &= \varepsilon \cos{(\alpha)}, \label{eq.4a} \\
    \epsilon_R &= \varepsilon \sin{(\alpha)} e^{i\psi}, \label{eq.4b}
\end{align}
\end{subequations}
where $\alpha$ controls the power distribution between the left- and right-handed modes, $\psi$ denotes their initial relative phase at the medium entrance, and $\varepsilon$ is an overall scaling factor.

\begin{figure}[!h]
    \centering
    \includegraphics[width=1\linewidth]{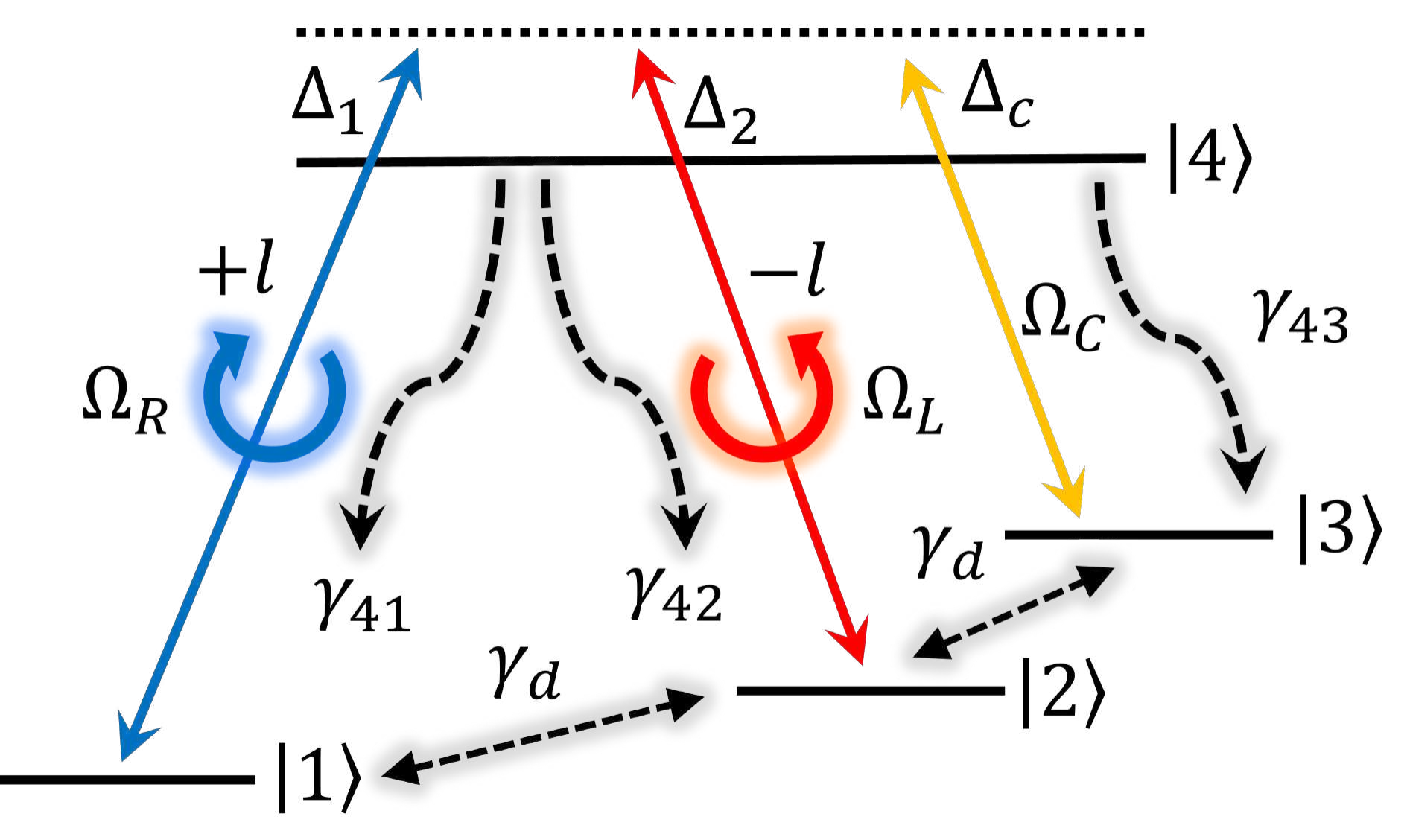}
    \caption{Schematic representation of the four-level atomic system in a tripod configuration. The excited state $\ket{4}$ is coupled to the ground states $\ket{1}$ and $\ket{2}$ by right- and left-circularly polarized probe fields with Rabi frequencies $\Omega_R$ and $\Omega_L$, respectively, and to the ground state $\ket{3}$ by a strong control field with Rabi frequency $\Omega_c$.}
    \label{Tripod slowlight configuration}
\end{figure}

The atomic medium is modeled as a four-level system comprising three ground states $\ket{1},\ket{2},$ and $\ket{3}$, a single excited state $\ket{4}$. The medium interacts weakly with the structured vector vortex probe field with an additional strong control field characterized by the Rabi frequency $\Omega_C$. The control field Rabi frequency is defined as $\Omega_{C}=\vec{d}_{C}\cdot \vec{E}_{C}/\hbar$, where $\vec{d}_{C}$ denotes the dipole moment of the atomic transition corresponding to the electric field component $\vec{E}_{C}$ \cite{scully.book1997}. The right-circularly polarized component of the probe field couples the $\ket{1}\rightarrow\ket{4}$ transition, while the left-circularly polarized component drives the
$\ket{2}\rightarrow\ket{4}$ transition. The control field couples the $\ket{3}\rightarrow\ket{4}$ transition. Together, these couplings constitute a tripod configuration, as illustrated in Fig.\ref{Tripod slowlight configuration}.

The practical implementation of this light-matter coupling scheme can be realized, for instance, in a system of $^{87}\mathrm{Rb}$ atomic vapours, where each of the bare-states involved in the configuration can be represented by the atom's Zeeman sub-levels \cite{lukin2001}. For instance, the coherently prepared ground-states $\ket{1}$ and $\ket{2}$ can be represented by the states $\ket{5S_{1/2},F=1,m=-1}$ and $\ket{5S_{1/2},F=1,m=+1}$, while the third unoccupied ground-state $\ket{3}$ can be represented by  $\ket{5S_{1/2},F=2,m=\pm1}$ and the excited state $\ket{4}$ is represented by $\ket{5P_{3/2},F=1,m=0}$. The OAM-less control field can either have the right- or left-handed circular polarization to couple the $\ket{5S_{1/2},F=2,m=+1}$ or $\ket{5S_{1/2},F=2,m=-1}$ with the excited state $\ket{5P_{3/2},F=1,m=0}$ according to dipole selection rules. To prepare the phaseonium state $\ket{1}$ and $\ket{2}$, an optical pumping can be implemented experimentally to induce the $\ket{5S_{1/2},F=1,m=0}\rightarrow \ket{5P_{3/2},F=0,m=0}$ transition. From the electric dipole selection rules and Clebsh-Gordan symmetry of $F=0\rightarrow F=1$ decay rates, the $\ket{5P_{3/2},F=0,m=0}$ population will decays equally to $m=-1,0,1$ in the $F=1$ manifold, thus effectively emptying $\ket{5S_{1/2},F=1,m=0}$ and  creating a dark-state from the superposition of $\ket{5S_{1/2},F=1,m=\pm 1}$ in this manifold \cite{Chen2006}. Moreover, all the vortex pairs and control field can be adjusted to propagate in the same direction to eliminate the Doppler effect, which degrades the EIT-related phenomena particularly in the Doppler-broadened medium \cite{Wang2018,Hsu:21}. 

Under the electric-dipole and rotating-wave approximations (RWA), the total Hamiltonian describing the atom–field interaction can be written as 
\begin{subequations}
\label{eq.5}
\begin{equation}
    H = H_0 + H_I, \label{eq.5a}
\end{equation}
\begin{equation}
    H_0 = \sum_{j=1}^4\hbar\omega_j\ket{j}\bra{j}, \label{eq.5b}
\end{equation}
\begin{multline}
    H_I = \hbar\Omega_R^*e^{-i\omega t}\ket{4}\bra{1}\\
    +\hbar\Omega_L^*e^{-i\omega t}\ket{4}\bra{2} + \hbar\Omega_C^*e^{-i\omega_C t}\ket{4}\bra{3} \\ + \mathrm{H.c.},\label{eq.5c}
\end{multline}
\end{subequations}  
where $\omega$ denotes the common carrier frequency of the right- and left-circularly polarized probe components, whereas $\omega_C$ represents the carrier frequency of the control field. The term $H_0$ corresponds to the unperturbed atomic Hamiltonian written in the bare-state basis, with eigenvalues $\hbar\omega_j$ ($j=1,2,3,4$). The operator $H_I$ describes the interaction between the atomic dipole moment and the applied optical fields within the electric-dipole approximation. 

In a rotating frame defined by the unitary transformation 
\begin{multline}
    W = e^{i(\omega_4 - \omega)t}\ket{1}\bra{1}+e^{i(\omega_4 - \omega)t}\ket{2}\bra{2} \\
    +e^{i(\omega_4-\omega_C)t}\ket{3}\bra{3}+e^{i\omega_4t}\ket{4}\bra{4}, \label{eq.6}
\end{multline}
and under the rotating-wave approximation, the total Hamiltonian takes the form  
\begin{equation}
    H' = i\hbar\frac{\partial W}{\partial t}W^\dagger + WHW^\dagger.\label{eq.7}
\end{equation}
Applying this transformation yields
\begin{multline}
        H' =-\hbar\Delta_1\ket{1}\bra{1} - \hbar\Delta_2\ket{2}\bra{2} -\hbar\Delta_C\ket{3}\bra{3}  \\ 
        + \hbar\Omega_R\ket{1}\bra{4} + \hbar\Omega_L\ket{2}\bra{4} + \hbar\Omega_C\ket{3}\bra{4} \\ 
        + H.c. , \label{eq.8}
\end{multline}
where the detunings of the optical fields are defined as
\begin{subequations}
\label{eq.9}
    \begin{align}
        \Delta_1  &= \omega_{41} - \omega;    \quad &\omega_{41} = \omega_4 - \omega_{1}, \label{eq.9a}\\
       \Delta_2 &= \omega_{42} - \omega; \quad &\omega_{42} = \omega_4 - \omega_{2}, \label{eq.9b} \\
       \Delta_C &= \omega_{43} - \omega_C; \quad &\omega_{43} = \omega_4 - \omega_{3}. \label{eq.9c} 
    \end{align}
\end{subequations}
\subsection{\label{Equation of Motion}Master Equations}
The dynamics of the atomic system are described by the evolution of the density matrix $\rho$ according to the Liouville–von Neumann equation
\begin{equation}
    \dot{\rho} = -\frac{i}{\hbar}\left[H,\rho\right] -\mathcal{L}\rho. \label{eq.10}
\end{equation}
The term $\mathcal{L}\rho$ represents relaxation processes, including both population decay and decoherence of the atomic coherences. In the present system, we consider two types of relaxation: radiative decay via spontaneous emission from the excited state, and non-radiative decoherence due to collisions among the ground states \cite{tarak2017}. Accordingly, the total relaxation operator is expressed as
\begin{equation}
    \label{eq.11}
    \mathcal{L}\rho = \mathcal{L}_r\rho + \mathcal{L}_c\rho,
\end{equation}
with the radiative relaxation term
\begin{equation}
    \label{eq.12}
    \mathcal{L}_r\rho=\sum_{j=1}^{3}\frac{\gamma_{4j}}{2}\left(\ket{4}\bra{4}\rho - \ket{j}\bra{j}\rho_{44} + \rho\ket{4}\bra{4} \right),
\end{equation}
and the non-radiative relaxation term
\begin{equation}
    \label{eq.13}
    \mathcal{L}_c\rho=\sum_{j=1}^{3}\sum_{j\neq i =1}^{3}\frac{\gamma_{d}}{2}\left(\ket{j}\bra{j}\rho - 2\ket{i}\bra{i}\rho_{jj} + \rho\ket{j}\bra{j} \right).
\end{equation}
Spontaneous emission induces relaxation from the excited state $\ket{4}$ to the ground states $\ket{j}$, with decay rates $\gamma_{4j}$ ($j=1,2,3$), while collisions among the ground states lead to dephasing of the coherences $\rho_{ij}$ ($i,j=1,2,3$) at a rate  $\gamma_d$. Substituting the time-independent Hamiltonian from Eq.~(\ref{eq.8}) together with the relaxation terms defined in Eqs.~(\ref{eq.11})-(\ref{eq.13}) into the master equation Eq.~(\ref{eq.10}) yields the following set of optical Bloch equations:
\begin{subequations}
\label{eq.14}
\begin{equation}
\begin{aligned}
\dot{\rho}_{11} = &i(\Omega_R^*\rho_{14}-\Omega_R\rho_{41}) + \gamma_{41}\rho_{44} \\ 
&+ \gamma_d(\rho_{22}+\rho_{33}-2\rho_{11}), \label{eq.14a} 
\end{aligned}
\end{equation}

\begin{equation}
\begin{aligned}
\dot{\rho}_{22} = &i(\Omega_L^*\rho_{24}-\Omega_L\rho_{42}) + \gamma_{42}\rho_{44} \\ &+ \gamma_d(\rho_{11}+\rho_{33}-2\rho_{22}), \label{eq.14b}
\end{aligned}
\end{equation}

\begin{equation}
\begin{aligned}
    \dot{\rho}_{33} = &i(\Omega_C^*\rho_{34}-\Omega_C\rho_{43}) + \gamma_{43}\rho_{44} \\ &+ \gamma_d(\rho_{11}+\rho_{22}-2\rho_{33}), \label{eq.14c}
\end{aligned}
\end{equation}

\begin{equation}
\begin{aligned}
    \dot{\rho}_{12} = &i(\rho_{12}(\Delta_1 - \Delta_2)+\Omega_L^*\rho_{14}-\Omega_L\rho_{41}) \\ &- 2 \gamma_d \rho_{12}, \label{eq.14d}
\end{aligned}
\end{equation}

\begin{equation}
\begin{aligned}
    \dot{\rho}_{13} = &i(\rho_{13}(\Delta_1 - \Delta_C) + \Omega_C^*\rho_{14}-\Omega_R\rho_{43}) \\ &- 2\gamma_d\rho_{13}, \label{eq.14e}
\end{aligned}
\end{equation}

\begin{equation}
\begin{aligned}
    \dot{\rho}_{14} = &i(\Omega_R \rho_{11} + \Omega_L \rho_{12} + \Omega_C \rho_{13} + \Delta_1\rho_{14})\\ &- i\Omega_R \rho_{44} - \Gamma\rho_{14} - \gamma_d\rho_{14}, \label{eq.14f}
\end{aligned}
\end{equation}

\begin{equation}
\begin{aligned}
    \dot{\rho}_{23} = &i(\rho_{23}(\Delta_2 - \Delta_C) + \Omega_C^*\rho_{24}-\Omega_L\rho_{43}) \\ &- 2\gamma_d\rho_{23}, \label{eq.14g}
\end{aligned}
\end{equation}

\begin{equation}
\begin{aligned}
    \dot{\rho}_{24} = &i(\Omega_R \rho_{21} + \Omega_L \rho_{22} + \Omega_C \rho_{23} + \Delta_2\rho_{24})\\ &- i\Omega_L \rho_{44} - \Gamma\rho_{24} - \gamma_d\rho_{24}, \label{eq.14h}
\end{aligned}
\end{equation}

\begin{equation}
\begin{aligned}
    \dot{\rho}_{34} = &i(\Omega_R \rho_{31} + \Omega_L \rho_{32} + \Omega_C \rho_{33} + \Delta_C\rho_{34})\\ &- i\Omega_C \rho_{44} - \Gamma\rho_{34} - \gamma_d\rho_{34}, \label{eq.14i}
\end{aligned}
\end{equation}
\end{subequations}
where $\Gamma=\frac{\gamma_{41}+\gamma_{42}+\gamma_{43}}{2}$. The remaining optical Bloch equations follow directly from the population conservation condition $\sum_{j=1}^4\rho_{jj}=1$ and the Hermitian property of the density matrix  $\dot{\rho}_{ij}=\dot{\rho}_{ji}^*$. 

We assume that the atomic ensemble is initially prepared in a coherent superposition of the ground states $\ket{1}$ and $\ket{2}$ commonly referred to as a phaseonium state. The initial state of the atoms, prior to interaction with the vortex probe fields and the control field, is expressed as
\begin{equation}
    \ket{\psi(0)} = \cos{(\theta)}\ket{1} + \sin{(\theta)}\ket{2}, \label{eq.15}
\end{equation}
where $\theta$ is a tunable angle that determines the relative population distribution between $\ket{1}$ and $\ket{2}$. We assume that both vortex components interact weakly with the medium, such that $|\Omega_R|,|\Omega_L|\ll|\Omega_C|$. Under these conditions, the steady-state coherences can be obtained using a perturbative expansion in the weak probe fields
\begin{equation}
    \rho_{ij} \approx \rho_{ij}^{(0)} + \rho_{ij}^{(1)} + \rho_{ij}^{(2)} + \rho_{ij}^{(3)} + \cdots, \label{eq.16} 
\end{equation}
where $\rho_{ij}^{(n)}$denotes the n-th order contribution in the probe fields. In the linear propagation regime, terms of second order and higher can be neglected. The zeroth-order term $\rho_{ij}^{(0)}$ is determined by the initial atomic state preparation given in Eq.~(\ref{eq.15}) and the assumption of weak atom–field interaction. In this limit, the excited state remains essentially unpopulated, yielding
\begin{subequations}
\label{eq.17}
\begin{align}
    \rho_{44}^{(0)} &= \rho_{33}^{(0)}\approx0, \label{eq.17a}\\
    \rho_{11}^{(0)}&\approx\cos^2{(\theta)}, \label{eq.17b}\\
    \rho_{22}^{(0)}&\approx\sin^2{(\theta)}, \label{eq.17c}\\
    \rho_{12}^{(0)}&=\rho_{21}^{(0)}\approx \sin{(\theta)}\cos{(\theta)}, \label{eq.17d}
\end{align}
\end{subequations}
while all remaining zeroth-order coherences $\rho_{ij}^{(0)}$ vanish. 

The first-order coherence terms $\rho_{14}^{(1)}$ and $\rho_{24}^{(1)}$ in the steady-state regime can be obtained by setting $\dot{\rho}_{ij}\approx0$ in Eq.~(\ref{eq.14}) and solving the resulting algebraic equations in Eqs.~(\ref{eq.14})-(\ref{eq.16}) using the zeroth-order elements from Eq.~(\ref{eq.17}). The first-order off-diagonal density matrix elements $\rho_{14}^{(1)}$ and $\rho_{24}^{(1)}$, which determine the linear susceptibility of the medium for the right- and left-handed components of the vortex field, are given by
\begin{subequations}
\label{eq.18}
\begin{align}
    \rho_{14}^{(1)} &= - \frac{\Omega_R\cos^2{(\theta)}+\Omega_L\sin{(\theta)}\cos{(\theta)}}{\Delta_1-\frac{|\Omega_C|^2}{\Delta_1-\Delta_C +2i\gamma_d} + i(\Gamma+\gamma_d)}, \label{eq.18a} \\
    \rho_{24}^{(1)} &= - \frac{\Omega_R\sin{(\theta)}\cos{(\theta)}+\Omega_L\sin^2{(\theta)}}{\Delta_2-\frac{|\Omega_C|^2}{\Delta_2-\Delta_C +2i\gamma_d} + i(\Gamma+\gamma_d)}, \label{eq.18b}
\end{align}
\end{subequations}
with $\rho_{14}^{(1)}=(\rho_{41}^{(1)})^*$ and $\rho_{24}^{(1)}=(\rho_{42}^{(1)})^*$. It should be noted that the first-order solutions in Eq.(\ref{eq.18}) are valid only under the condition $|\rho_{41}^{(1)}|,|\rho_{42}^{(1)}|\ll1$. When both probe components interact strongly with the atomic medium, the populations of the ground states $\rho_{11}$ and $\rho_{22}$ evolve significantly during propagation, and the full time dependence of all density matrix elements must be taken into account. In the weak-probe limit, the excited-state population $\rho_{44}$ does not contribute at first order; it appears only in higher-order perturbative terms (second order or above) relevant to the nonlinear propagation regime. These higher-order contributions to $\rho_{44}$ induce optical pumping from the ground states $\ket{1}$ and $\ket{2}$, leading to a depletion of the initially prepared coherent superposition given in Eq.~(\ref{eq.15}), as the excited-state population can decay incoherently. 

\subsection{\label{MBE}Field Propagation Equations}
The propagation of a structured optical field through a tripod atomic medium can be described by coupled Maxwell–Bloch equations for the right- and left-circularly polarized components of the field. Under the slowly varying envelope and paraxial approximations, the evolution of the Rabi frequencies $\Omega_R$ and $\Omega_L$ is governed by \cite{Hamid2023}
\begin{subequations}
\label{eq.19}
    \begin{equation}
        \frac{\partial \Omega_R}{\partial z} + c^{-1} \frac{\partial \Omega_R}{\partial t} = i\zeta\rho_{14}, \label{eq.19a}
    \end{equation}
    \begin{equation}
        \frac{\partial \Omega_L}{\partial z} + c^{-1} \frac{\partial \Omega_L}{\partial t} = i \zeta\rho_{24}, \label{eq.19b}
    \end{equation}
\end{subequations}
where $\zeta=\frac{2\pi N |d|^2 \omega}{c}$, with $c$ the speed of light in vacuum, $N$ the atomic number density, and $|d|$ the dipole moment, assumed equal for the transitions driven by the right- and left-handed vortex components ($d_R=d_L=d$). The transverse derivatives corresponding to diffraction effects have been neglected in Eq.~(\ref{eq.19}) \cite{Hamid.PRA2019,Hamid2023,Kudriasov.OE2025,Permana2025}.

To simplify the analysis, we assume that the two vortex fields have equal detuning, $\Delta_1=\Delta_2=\Delta$. Under this assumption, and by combining Eq.~(\ref{eq.18}) and (\ref{eq.19}) in the steady-state regime, the evolution of the vortex fields can be expressed as
\begin{equation}
    \frac{\partial}{\partial z}
    \begin{bmatrix}
        \Omega_R \\
        \Omega_L
    \end{bmatrix}
    = -iK
    \begin{bmatrix}
    \Omega_R \\
    \Omega_L
    \end{bmatrix}, \label{eq.20}
\end{equation}
where $K$ is a matrix defined by
\begin{equation}
    K = Q
    \begin{bmatrix}
        \cos^2{(\theta)} & \sin{(\theta)}\cos{(\theta)} \\
        \sin{(\theta)}\cos{(\theta)} & \sin^2{(\theta)}
    \end{bmatrix}, \label{eq.21}
\end{equation}
with $Q$ satisfying
\begin{equation}
    Q = \zeta \frac{1}{\Delta- \frac{|\Omega_C|^2}{\Delta-\Delta_C + 2i\gamma_d} + i(\Gamma+\gamma_d)}, \label{eq.22}
\end{equation}
which depends on the atomic parameters $\zeta$, the two-photon detuning $\Delta$, the control field detuning $\Delta_c$, the radiative decay rate $\Gamma$ and non-radiative dephasing rate $\gamma_d$.

The solution of Eqs.~(\ref{eq.20})-(\ref{eq.22}) depends on the initial conditions at the entrance of the medium $z=0$, denoted by $\Omega_R(0)=\Omega_{R,0}$ and $\Omega_L(0)=\Omega_{L,0}$. The corresponding solutions for the Rabi frequencies $\Omega_R(z)$ and $\Omega_L(z)$ along the propagation direction are then given by 
\begin{subequations}
\label{eq.23}
\begin{equation}
\begin{aligned}
    \Omega_R(z) = &\Omega_{R,0}\left(\sin^2{(\theta)} 
     +\cos^2{(\theta)}e^{-iQz}\right) \\ 
     &+ \Omega_{L,0} \sin{(\theta)}\cos{(\theta)} \left(e^{-iQz} - 1 \right), \label{eq.23a}
\end{aligned}
\end{equation}

\begin{equation}
\begin{aligned}
    \Omega_L(z) = &\Omega_{R,0}\sin{(\theta)}\cos{(\theta)} \left(e^{-iQz} - 1 \right) \\ 
     &+ \Omega_{L,0}  \left(\cos^2{(\theta)} 
     +\sin^2{(\theta)}e^{-iQz}\right), \label{eq.23b}
\end{aligned}
\end{equation}
\end{subequations}
where $\Omega_{R,0}$ and $\Omega_{L,0}$ are determined according to the definitions in Eqs.~(\ref{eq.2})-(\ref{eq.4}). 

The evolution of the right-handed component $\Omega_R(z)$ and the left-handed component $\Omega_L(z)$ at any position $z$ inside the atomic medium can be analyzed using Eq.~(\ref{eq.23}). This equation demonstrates that the initially prepared right- and left-handed vortex components at the medium entrance $z=0$ are mutually coupled through their interaction with the tripod atomic ensemble, resulting in a propagation behavior distinct from free-space evolution. In general, both $\Omega_R(z)$ and $\Omega_L(z)$ experience attenuation upon entering the medium due to the imaginary part of $Q$ defined in Eq.~(\ref{eq.22}). The detailed effects of this attenuation and absorption will be discussed in Section \ref{absorption}. Furthermore, in the case where only one component is initially present at $z=0$, Eq.~(\ref{eq.23}) reduces to a special case of slow-light vortex transfer, as demonstrated in \cite{Hamid2023}, highlighting the generality of this solution.
    
\subsection{\label{Susceptibility}Polarization-Dependent Susceptibilities}
The linear susceptibility of the medium for the right- and left-circularly polarized components can be expressed as \cite{Marangos.RMP2005}:
\begin{subequations}
\label{eq.24}
\begin{align}
    \chi_R^{(1)} &= -\frac{4\pi N|d|^2}{\Omega_R} \rho_{14}^{(1)}, \label{eq.24a} \\
    \chi_L^{(1)} &= -\frac{4\pi N|d|^2}{\Omega_L} \rho_{24}^{(1)}. \label{eq.24b}
\end{align}
\end{subequations}
By substituting the first-order coherence terms from Eq.~(\ref{eq.18}) under the assumption of equal detuning for both components, the susceptibilities can be rewritten as  
\begin{subequations}
\label{eq.25}
\begin{align}
    \chi_R^{(1)} &= \frac{2\zeta c}{\omega} \frac{\cos^2{(\theta)}+\frac{\Omega_L}{\Omega_R}\sin{(\theta)}\cos{(\theta)}}{\Delta-\frac{|\Omega_C|^2}{\Delta-\Delta_C +2i\gamma_d} + i(\Gamma+\gamma_d)}, \label{eq.25a} \\
    \chi_L^{(1)} &= \frac{2\zeta c}{\omega}\frac{\frac{\Omega_R}{\Omega_L}\sin{(\theta)}\cos{(\theta)}+\sin^2{(\theta)}}{\Delta-\frac{|\Omega_C|^2}{\Delta-\Delta_C +2i\gamma_d} + i(\Gamma+\gamma_d)}, \label{eq.25b}
\end{align}
\end{subequations}
where $\Omega_{R(L)}$ are taken from the propagation solution in Eq.~(\ref{eq.23}). In general, Eq.~(\ref{eq.25}) shows that the medium’s susceptibility differs for the two polarization components, and is controlled by the initial atomic coherence angle $\theta$, as well as the topological charge $l$ of the vortex, appearing in the phase-dependent terms $\frac{\Omega_{R(L)}}{\Omega_{L(R)}}\sim \mathrm{exp}(\pm2il\phi)$. This phase-dependent anisotropy affects both the polarization state and the intensity profile of the beam during propagation, as will be discussed in detail in Sections \ref{absorption} and \ref{propagation}.

\section{Propagation of Structured Light}
\subsection{\label{absorption}Absorption and Dispersion Characteristics}
We first examine the absorption and dispersion characteristics of the right- and left-circularly polarized beam components. The absorption is determined by the imaginary part of the susceptibility Im$[\chi_{R(L)}^{(1)}]$, while the dispersion is governed by the real part, Re$[\chi_{R(L)}^{(1)}]$ as given in Eq.~(\ref{eq.25}). We assume that the non-radiative dephasing rate $\gamma_d$ is much smaller than the radiative decay rate $\Gamma$ with $\gamma_d\sim10^{-3}\Gamma$, which is a reasonable approximation for cold atomic systems \cite{Rebic2004,tarak2017}.

\begin{figure*}[!t]
\centering
\includegraphics[width=1\linewidth,height=0.7\textheight]{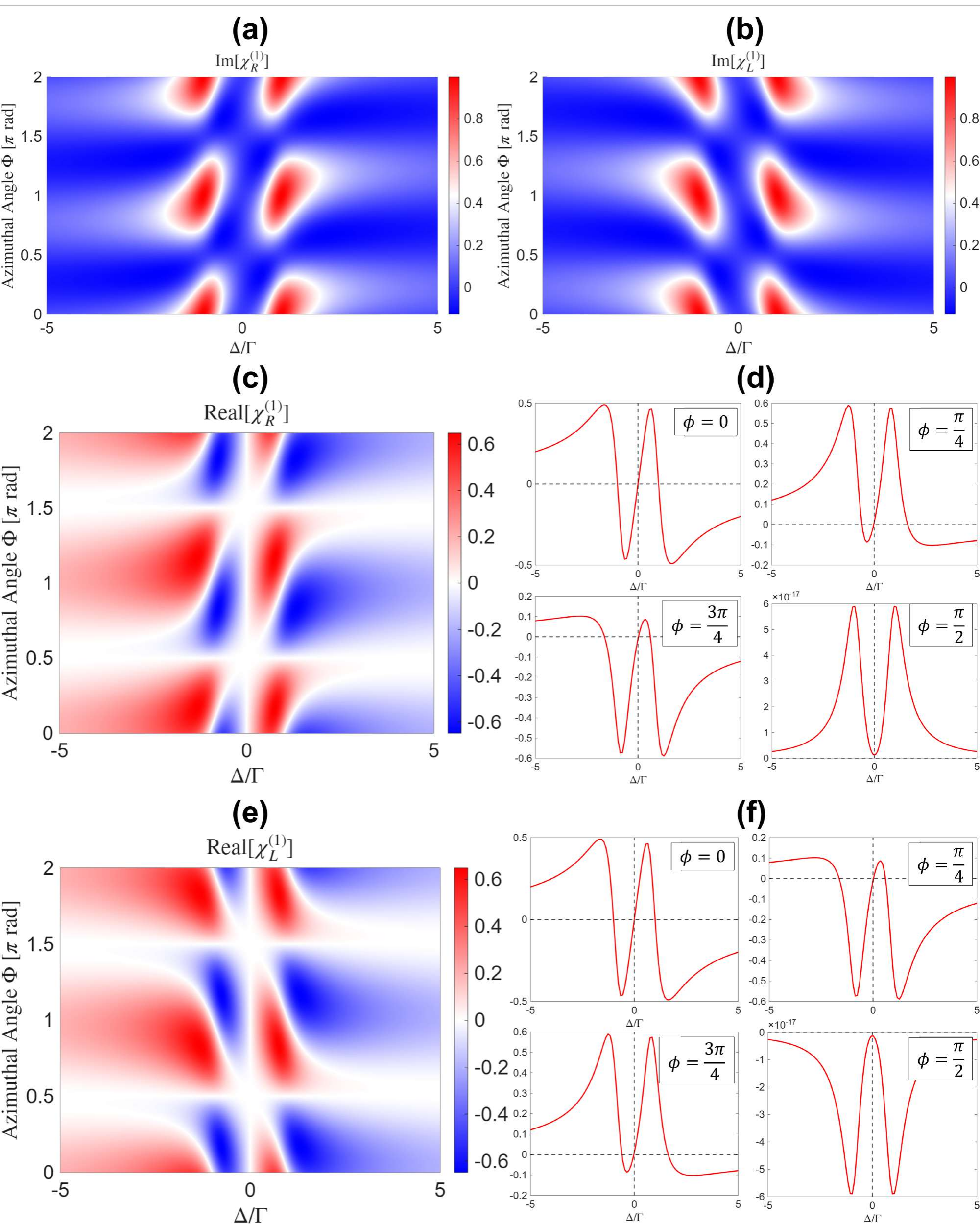}
    \caption{Absorption (Im[$\chi^{(1)}$]) (a)–(b) and dispersion (Re[$\chi^{(1)}$]) (c)–(f) profiles for the right- and left-handed components of the vector vortex beam with topological charge $|l|=1$ at $z=0$. Panels (a) and (b) show the imaginary part of the susceptibilities evaluated at $r=w$ as functions of azimuthal angle $\phi$ and detuning $\Delta/\Gamma$ for the right- and left-handed components, respectively. Panels (c) and (e) present the corresponding real parts. Panels (d) and (f) display the one-dimensional dispersion profiles, Re[$\chi_R^{(1)}$] and Re[$\chi_L^{(1)}$], plotted versus $\Delta/\Gamma$ for selected azimuthal angles $\phi = 0, \pi/4, \pi/2, 3\pi/4$. In (d) and (f), vertical dashed lines indicate resonance ($\Delta/\Gamma = 0$), while horizontal dashed lines denote Re[$\chi_{R(L)}^{(1)}$] = 0. The parameters used are $\gamma_d = 10^{-3}\Gamma$, $|\Omega_C| = \Gamma$, $\theta = \pi/4$, $\alpha = \pi/4$, and $\psi = 0$. All plots are normalized to $2\frac{c}{\omega}\zeta$.}
    \label{Absorption at z=0}
\end{figure*}

For simplicity, we consider the symmetric case in which the two ground states $\ket{1}$ and $\ket{2}$ are equally populated by setting the coherence tuning angle to $\theta=\pi/4$. We further assume that, at the entrance of the medium ($z=0$), the right- and left-circularly polarized components possess equal amplitudes and no relative phase difference. This condition is imposed by choosing $\alpha=\pi/4$ and $\psi=0$ in Eq.~(\ref{eq.4}). Under these assumptions, the initial field amplitudes satisfy $|\Omega_{R,0}|=|\Omega_{L,0}|=|\tilde{\Omega}_0|=\varepsilon A(r)$, where $A(r)$ is defined in Eq.~(\ref{eq.3}). With these symmetric initial conditions, the propagation solutions given in Eq.~(\ref{eq.23}) simplify to
\begin{subequations}
\label{eq.26}   
\begin{align}
    \Omega_R(z) = &\varepsilon A(r)\left(e^{-iQz}\cos{(l\phi)}+i\sin{(l\phi)} \right), \label{eq.26a} \\
    \Omega_L(z) = &\varepsilon A(r)\left(e^{-iQz}\cos{(l\phi)} - i\sin{(l\phi)} \right). \label{eq.26b}
\end{align}
\end{subequations}
Under the same conditions, the susceptibilities in Eq.~(\ref{eq.25}) reduce to
\begin{subequations}
\label{eq.27}
\begin{align}
    \chi_R^{(1)} &= \frac{2c}{\omega}Q \frac{e^{-iQz}}{e^{-iQz}+i\tan{(l\phi)}}
     \label{eq.27a} \\
    \chi_L^{(1)} &= \frac{2c}{\omega}Q \frac{e^{-iQz}}{e^{-iQz}-i\tan{(l\phi)}}. \label{eq.27b}
\end{align}
\end{subequations} 



It can be seen from Eqs.~(\ref{eq.27a}) and (\ref{eq.27b}) that the susceptibilities, and consequently the absorption and dispersion characteristics, are polarization dependent. This implies that the different components of the optical vector vortex beam experience distinct attenuation and group delay during propagation inside the atomic medium. The origin of this behavior lies in the coherent superposition of $\ket{1}$ and $\ket{2}$ defined in Eq.~(\ref{eq.15}), which induces an effective anisotropy in the medium, even in the absence of the control field. When the control field is absent, azimuthally dependent transparency windows emerge at resonance $\Delta=0$ with a $2|l|$ degeneracy as shown recently in \cite{Permana2025}. In contrast, when a non-zero control field is introduced, a uniform transparency window forms around $\Delta=0$, allowing the entire transverse profile of the beam to propagate without absorption. In this case, the dispersion exhibits a positive slope in the vicinity of resonance, leading to slow-light propagation of the vector vortex beam, as will be demonstrated later.

To illustrate the absorption and dispersion behavior in the presence of the control field, we assume that the control field is resonant with the $\ket{1} \rightarrow \ket{4}$ transition, i.e., $\Delta_C = 0$. Under this condition, the absorption and dispersion properties are primarily governed by the two-photon detuning $\Delta$. Figure \ref{Absorption at z=0} presents the susceptibility characteristics of the vector vortex beam with $|l|=1$ at the entrance of the medium $z=0$. Panels (a) and (b) depict the absorption profiles of the right- and left-handed components, respectively. The corresponding dispersion behavior is shown in panels (c) and (d) for the right-handed component, and in panels (e) and (f) for the left-handed component. The control field amplitude is fixed at $|\Omega_C|=\Gamma$. All quantities are normalized by $\frac{2c}{\omega}\zeta$, ensuring that the results can be readily extended to different cold atomic gas media. 

It can be observed from all subfigures in Fig. \ref{Absorption at z=0} that both the absorption and dispersion exhibit a pronounced phase dependence, as their profiles vary with the azimuthal angle $\phi$. This phase-dependent behavior originates from the $e^{-2il\phi}$ term appearing in the susceptibility of the right-handed component and the $e^{+2il\phi}$ term in that of the left-handed component. Consequently, the absorption and dispersion of both polarization components vary periodically with $\phi$, with a periodicity of $\frac{\pi}{|l|}$. For example, in the case $|l|=1$ shown in Fig. \ref{Absorption at z=0}, the periodicity reduces to $\pi$. This can be clearly seen in subplots (a)–(d), where the absorption and dispersion profiles of both the right- and left-handed components repeat for successive increments of $\phi=\pi$.

The imaginary parts of $\chi_R^{(1)}$ and $\chi_L^{(1)}$ exhibit a rich structure consisting of absorption, transparency, and gain regions. From subfigures (a) and (b) in Fig. \ref{Absorption at z=0}, it can be seen that near resonance ($\Delta/\Gamma = 0$), both the right- and left-handed components experience EIT (indicated by the blue color corresponding to $\mathrm{Im}[\chi_{R(L)}^{(1)}] = 0$). This transparency extends across the entire azimuthal plane. Away from exact resonance, however, the response becomes azimuthally modulated. Two absorption maxima emerge around $|\Delta| = \Gamma$, shown in red and corresponding to $\mathrm{Im}[\chi_{R(L)}^{(1)}] > 0$, indicating enhanced attenuation at specific angular positions. In addition, small gain regions are observed, represented by darker blue areas where $\mathrm{Im}[\chi_{R(L)}^{(1)}] < 0$.

On the other hand, the dispersion profiles of the right-handed (Fig. \ref{Absorption at z=0}(c)) and left-handed (Fig. \ref{Absorption at z=0}(e)) components show that, in the vicinity of resonance $|\Delta|<0.5\Gamma$, the real part of the susceptibility transitions from negative to positive as the detuning increases. This behavior corresponds to a positive slope of dispersion near $\Delta = 0$. A positive dispersion slope implies a positive group index, leading to subluminal (slow-light) propagation \cite{Marangos.RMP2005}. This feature becomes more evident in the one-dimensional dispersion profiles shown in Fig. \ref{Absorption at z=0}(d) for $\mathrm{Re}[\chi_{R}^{(1)}]$ and Fig. \ref{Absorption at z=0}(f) for $\mathrm{Re}[\chi_{L}^{(1)}]$, where specific azimuthal angles are selected. These plots confirm that, for most values of $\phi$, both circular polarization components exhibit a positive dispersion slope at resonance (indicated by the vertical dashed lines at $\Delta/\Gamma = 0$), demonstrating that slow-light propagation dominates across the transverse plane. Exceptions occur at $\phi = \pi/2$ and $\phi = 3\pi/2$, where the slope of the dispersion vanishes, corresponding to a zero group index at those specific angular positions. Nevertheless, apart from these isolated angles, the vector vortex beam propagates predominantly in the slow-light regime due to the presence of the control field, which establishes the EIT window. This reduced group-velocity behavior is absent in previous studies of vector vortices in EIT systems without a sufficiently strong control field \cite{radwell2017,tarak2017,Kudriasov.OE2025,Permana2025}. The ability to sustain slow-light propagation in the structured beam therefore represents an important feature for investigating vector vortex dynamics in the subluminal regime.

\begin{figure*}[!t]
  \centering
  \includegraphics[width=0.49\textwidth]{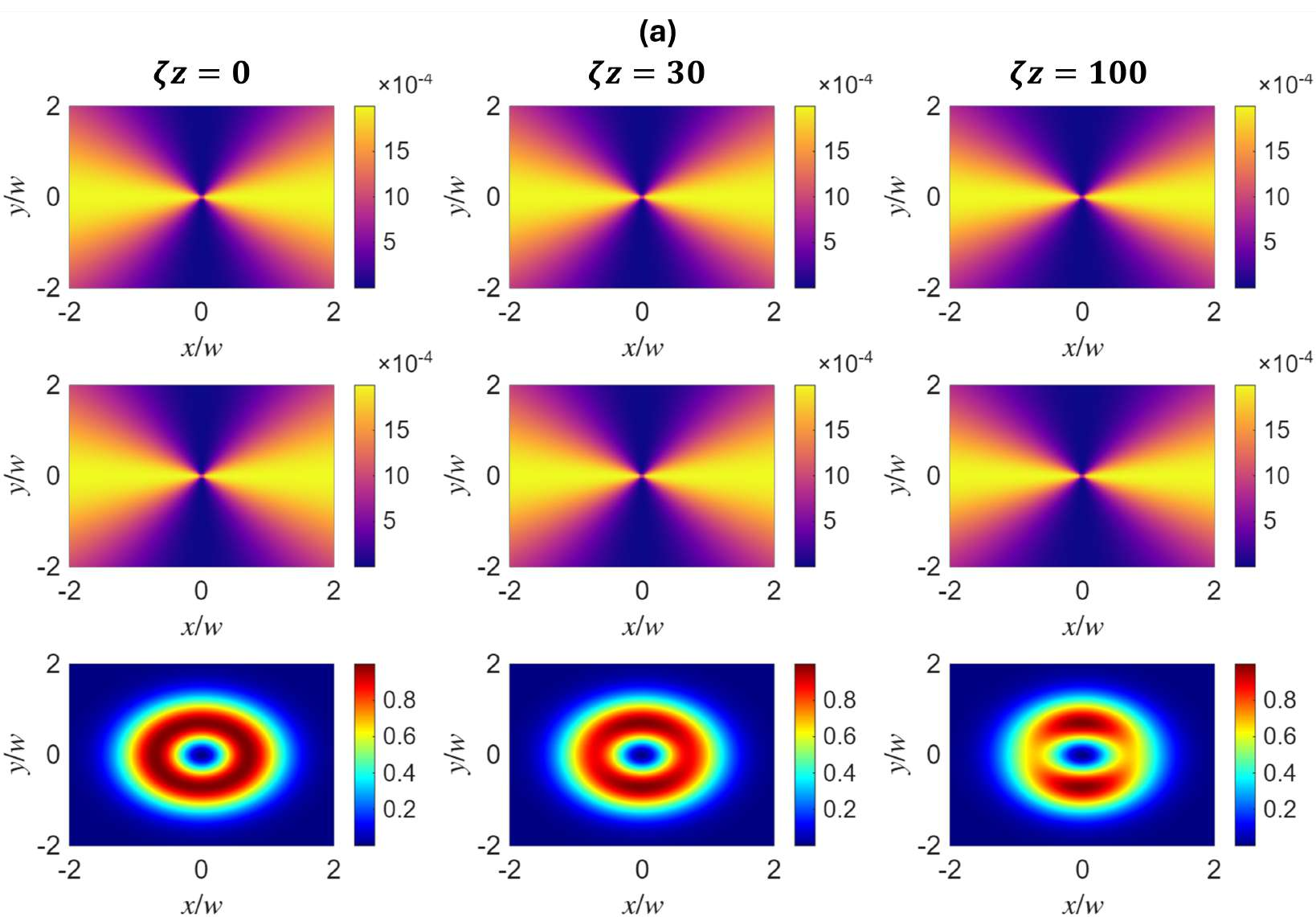}
  \hfill
  \includegraphics[width=0.49\textwidth]{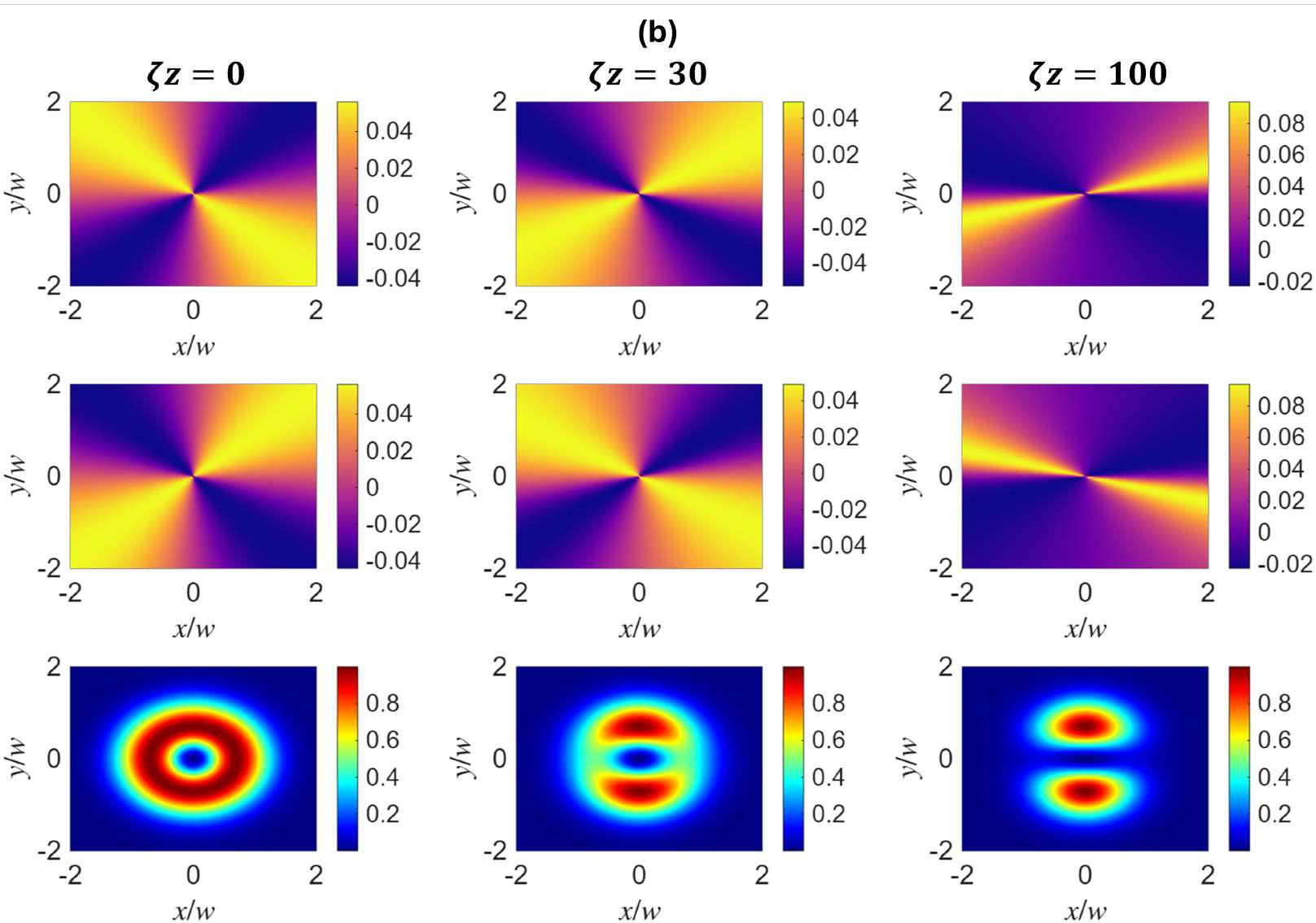}
  \caption{Evolution of the transverse absorption profiles of the right- and left-handed components, together with the total intensity distribution, at different normalized propagation distances $\zeta z$ for $|l|=1$. The first row shows the absorption of the right-handed component, the second row shows the absorption of the left-handed component, and the third row shows the combined total intensity of both components. (a) Two-photon resonance $\Delta=0$. (b) Slight detuning from resonance with $\Delta = 0.1\Gamma$. All other parameters are the same as in Fig. \ref{Absorption at z=0}.
  }
  \label{Beam Intensity and Susceptibility Evolution}
\end{figure*}

Figure \ref{Beam Intensity and Susceptibility Evolution}(a) illustrates the evolution of the absorption profiles of the right- and left-handed components for $|l|=1$, together with the corresponding total intensity distribution in the transverse plane, at different normalized propagation distances $\zeta z$ under exact two-photon resonance $\Delta=0$. The first and second rows display the absorption patterns of the right- and left-handed components, respectively, while the third row shows the evolution of the total intensity. For $|l|=1$, two azimuthal transparency windows are formed, appearing as dark regions corresponding to $\mathrm{Im}[\chi_{R(L)}^{(1)}]=0$. At the same time, two complementary regions of non-zero absorption (bright areas) are present, although with reduced magnitude compared to the no-control-field case \cite{Permana2025}. This indicates that the $2|l|$-fold degeneracy remains preserved in the presence of the control field, but with suppressed absorption strength. As a consequence, the transformation of the beam profile from an initial ring-shaped intensity distribution into a petal-like structure occurs more gradually, requiring a longer propagation distance for the petals to fully develop, as observed in the last row.

Figure \ref{Beam Intensity and Susceptibility Evolution}(b) shows the corresponding evolution under slight detuning ($\Delta=0.1\Gamma$). In this regime, the right- and left-handed components exhibit complementary spatial patterns of amplification and absorption. In the first row, corresponding to the right-handed component, two amplification windows appear for $|l|=1$, shown as dark regions where $\mathrm{Im}[\chi_R^{(1)}]<0$. These amplification regions spatially coincide with absorption windows of the left-handed component (second row), where $\mathrm{Im}[\chi_L^{(1)}]>0$, and vice versa. This complementary gain–loss behavior originates from the structure of the susceptibilities in Eq.~(\ref{eq.25}). The right-handed susceptibility contains the ratio $\Omega_L/\Omega_R$, producing the phase factor $\exp(-2il\phi)$, whereas the left-handed susceptibility contains $\Omega_R/\Omega_L$, yielding $\exp(2il\phi)$. These azimuthal phase terms $\exp(\pm2il\phi)$ generate spatially dependent gain and loss for non-zero OAM charge $l$, with opposite rotational symmetry for the two polarization components. Despite the opposite local gain–loss dynamics, the total intensity of the beam still evolves from a ring into a petal-like structure. However, each polarization component undergoes a dynamic propagation process in which the gain–loss regions rotate and change in size. This evolving spatial imbalance leads to a continuously varying local polarization state, which will be analyzed in the next section.

\subsection{\label{propagation}Polarization dynamics of slow-light vector vortices}

In this section, we analyze in detail the polarization evolution of slow-light vector vortices as they propagate inside the tripod atomic medium. The total electric field at an arbitrary propagation distance $z$ is expressed as the superposition of its circular polarization components, $\vec{E}(r,\phi,z)=E_R(z)\vec{e}_R + E_L(z)\vec{e}_L$, where $E_R(z)$ and $E_L(z)$ are obtained from the solutions in Eq.~(\ref{eq.23}). The relation between the Rabi frequencies and the corresponding electric field amplitudes is given by $\Omega_{R(L)} = \vec{d}\cdot\vec{E}_{R(L)}/\hbar$. Assuming equal dipole moments for the right- and left-handed transitions ($d_R = d_L = d$), the field amplitudes follow directly from the Rabi solutions. Explicitly, the circular components of the optical vector vortex can be written as
\begin{subequations}
\label{eq.28}
\begin{equation}
\begin{aligned}
    E_R(z) = &\tilde{\varepsilon} \sin{(\alpha)}e^{i\psi}A(r)e^{il\phi}\left(\sin^2{(\theta)} 
     +\cos^2{(\theta)}e^{-iQz}\right) \\ 
     &+ \tilde{\varepsilon} \cos{(\alpha)}A(r)e^{-il\phi} \sin{(\theta)}\cos{(\theta)} \left(e^{-iQz} - 1 \right), \label{eq.28a}
\end{aligned}
\end{equation}

\begin{equation}
\begin{aligned}
    E_L(z) = &\tilde{\varepsilon} \sin{(\alpha)}e^{i\psi}A(r)e^{il\phi}\sin{(\theta)}\cos{(\theta)} \left(e^{-iQz} - 1 \right) \\ 
     &+ \tilde{\varepsilon} \cos{(\alpha)}A(r)e^{-il\phi}  \left(\cos^2{(\theta)} 
     +\sin^2{(\theta)}e^{-iQz}\right), \label{eq.28b}
\end{aligned}
\end{equation}
\end{subequations}
where $A(r)$ is defined in Eq. (\ref{eq.3}), and $\tilde{\varepsilon}$ is a scaling factor related to the Rabi amplitude $\varepsilon$ through $\tilde{\varepsilon}=\hbar\varepsilon/d$. By tuning the parameters $\alpha$ and $\theta$, the input polarization structure and the initial atomic population distribution can be independently controlled, shaping the subsequent polarization evolution. 

\begin{figure*}[!t]
    \centering
    \includegraphics[width=0.48\linewidth]{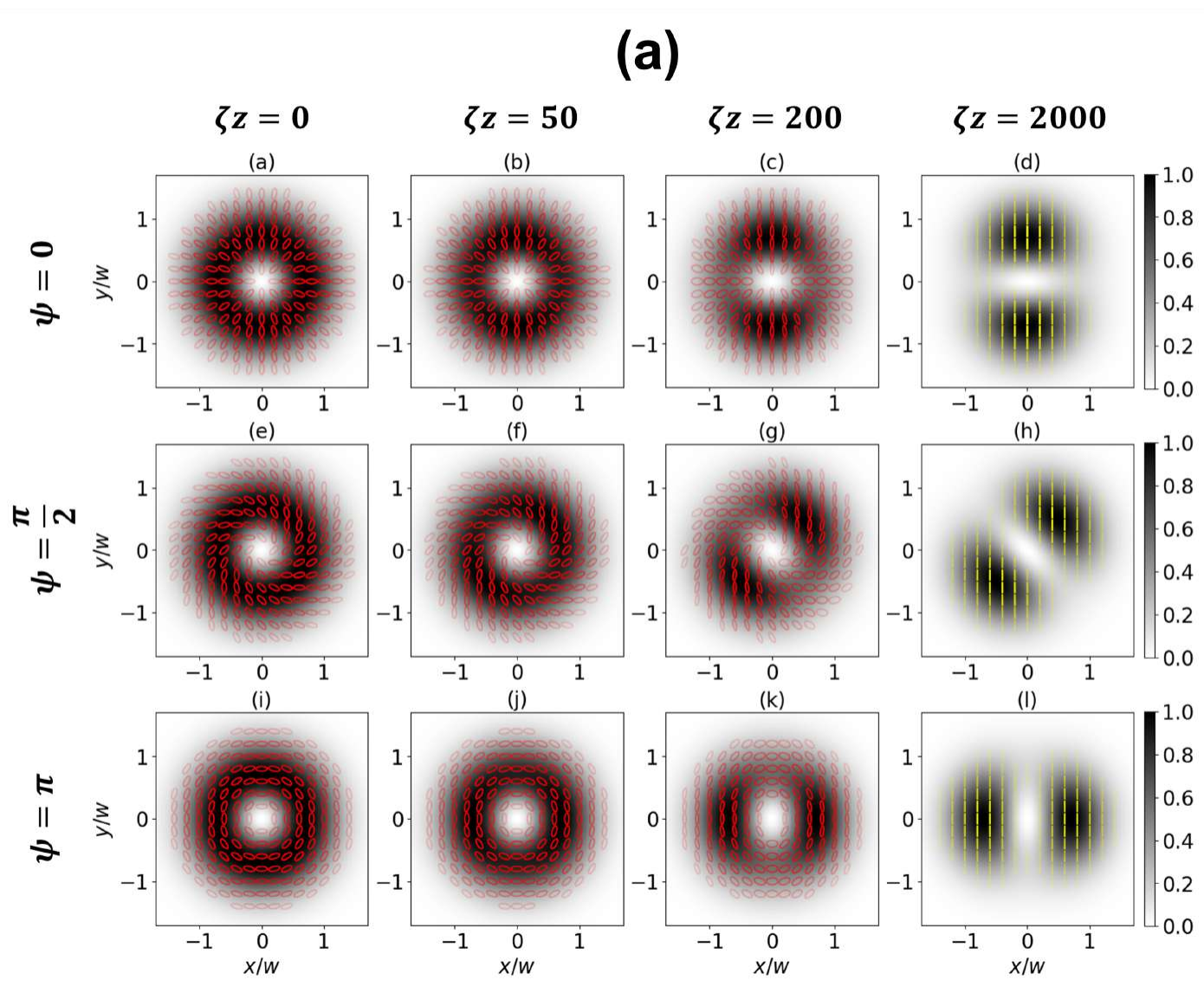}
    \hfill
    \includegraphics[width=0.48\linewidth]{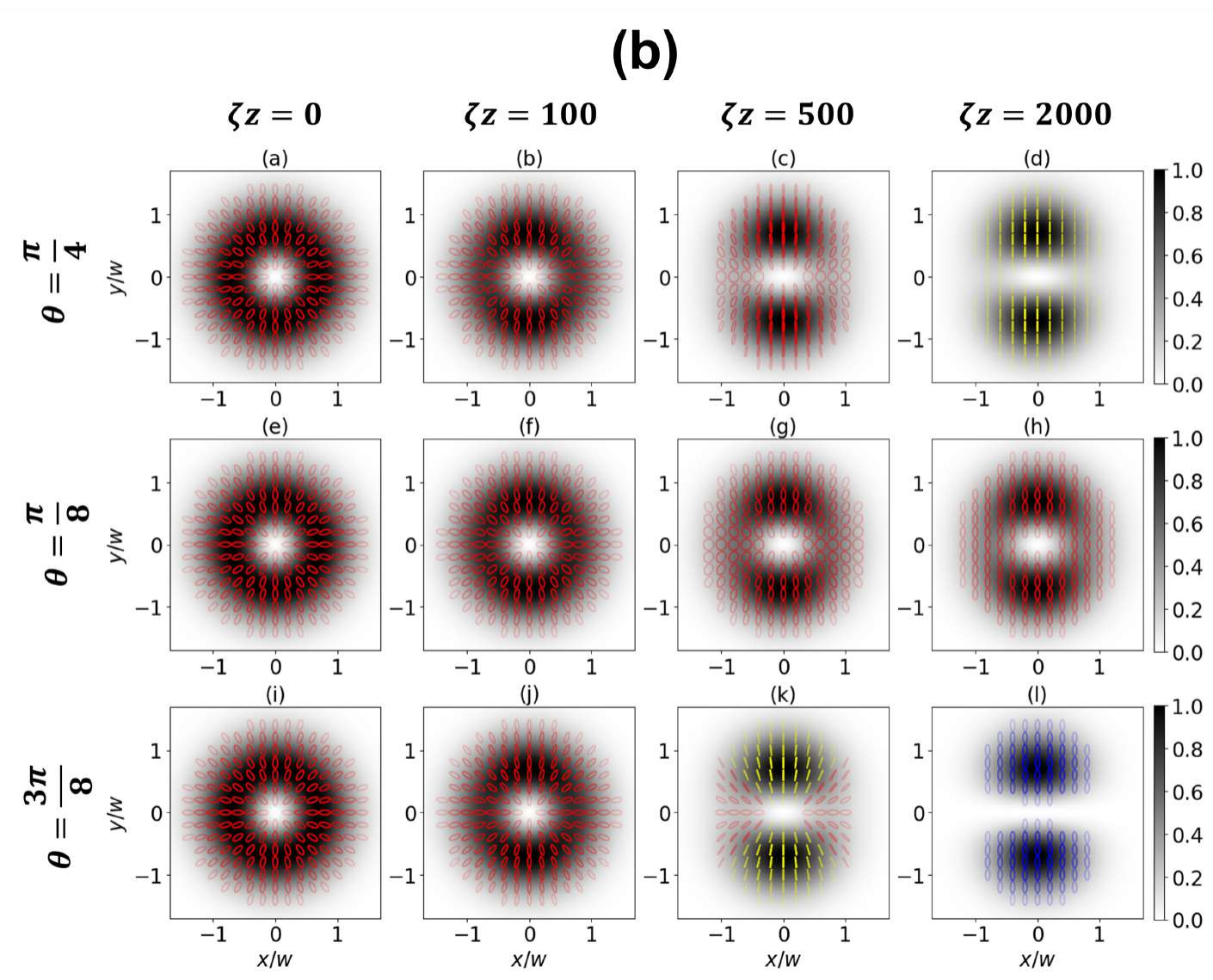}
    \caption{Intensity profile and polarization state evolution in the transverse plane of the vector vortex with $|l|=1$ and relative amplitude $\alpha=\pi/8$, such that the initial polarization state at the medium entrance is dominated by the left-handed circular polarization ($|E_L(0)|>|E_R(0)|$). The transverse coordinates ($x,y$) are normalized to the beam waist $w$, with the propagation distance expressed as the dimensionless parameter $\zeta z$. The vector vortex beam is in two-photon resonance with detuning $\Delta=0$, while the other parameters are $\gamma_d=10^{-3}\Gamma$ and $|\Omega_C|=\Gamma$. Darker rings or lobes correspond to higher intensity regions. The red and blue ellipses denote the  the left- and right-circular polarization states, respectlively, while yellow lines indicate the linear polarization state. (a) The initial phaseonium state is fixed at $\theta=\pi/4$; the first, second, and third rows correspond to relative phases $\psi=0,\pi/2,\pi$, respectively. (b) The relative phase is fixed at $\psi=0$; the first, second, and third rows correspond to initial phaseonium state $\theta=\pi,\pi/4,3\pi/8$, respectively.}
    \label{Polarization Fixed Coherence}
\end{figure*}

\begin{figure*}[!t]
    \centering
    \includegraphics[width=0.48\linewidth]{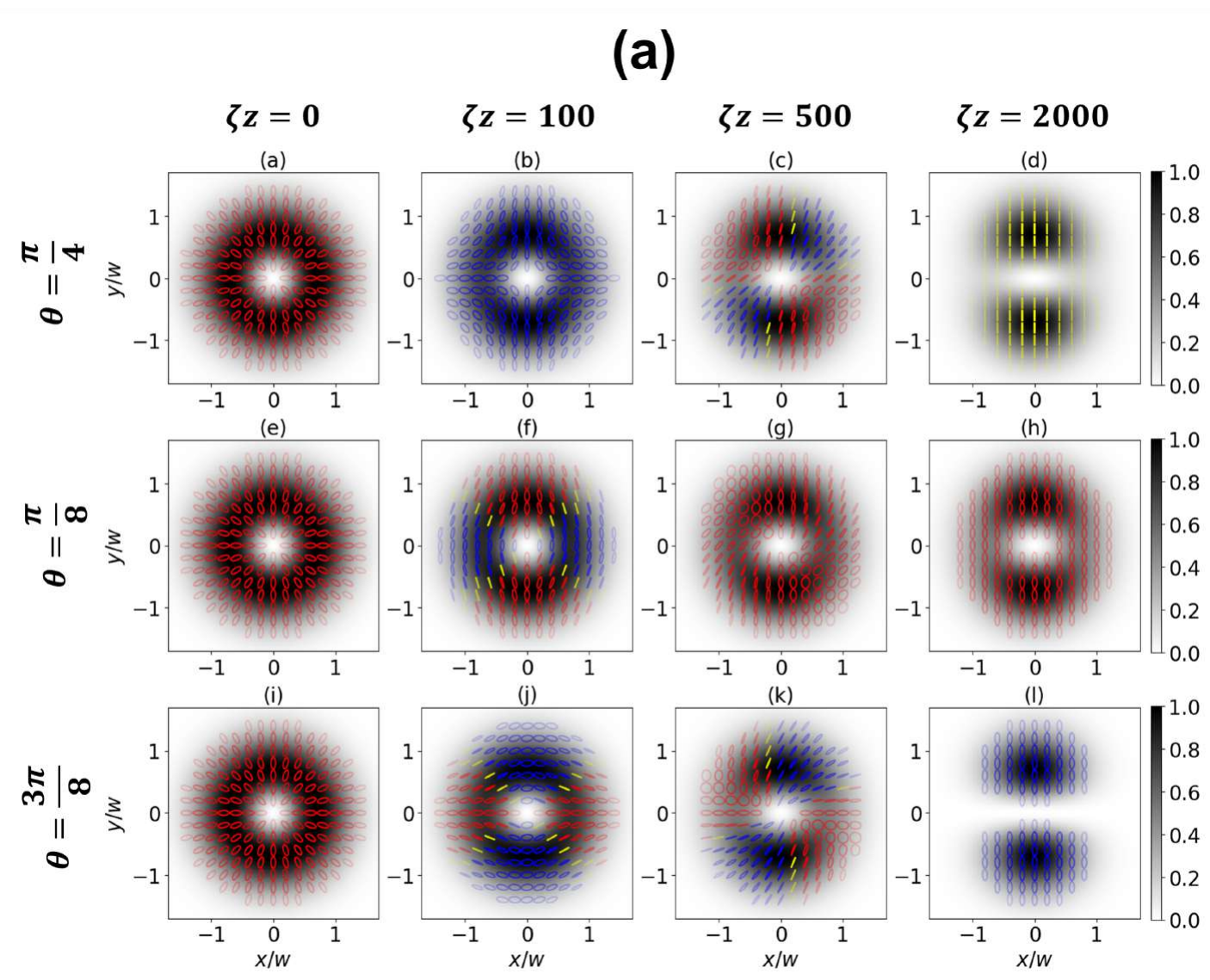}
    \hfill
    \includegraphics[width=0.48\linewidth]{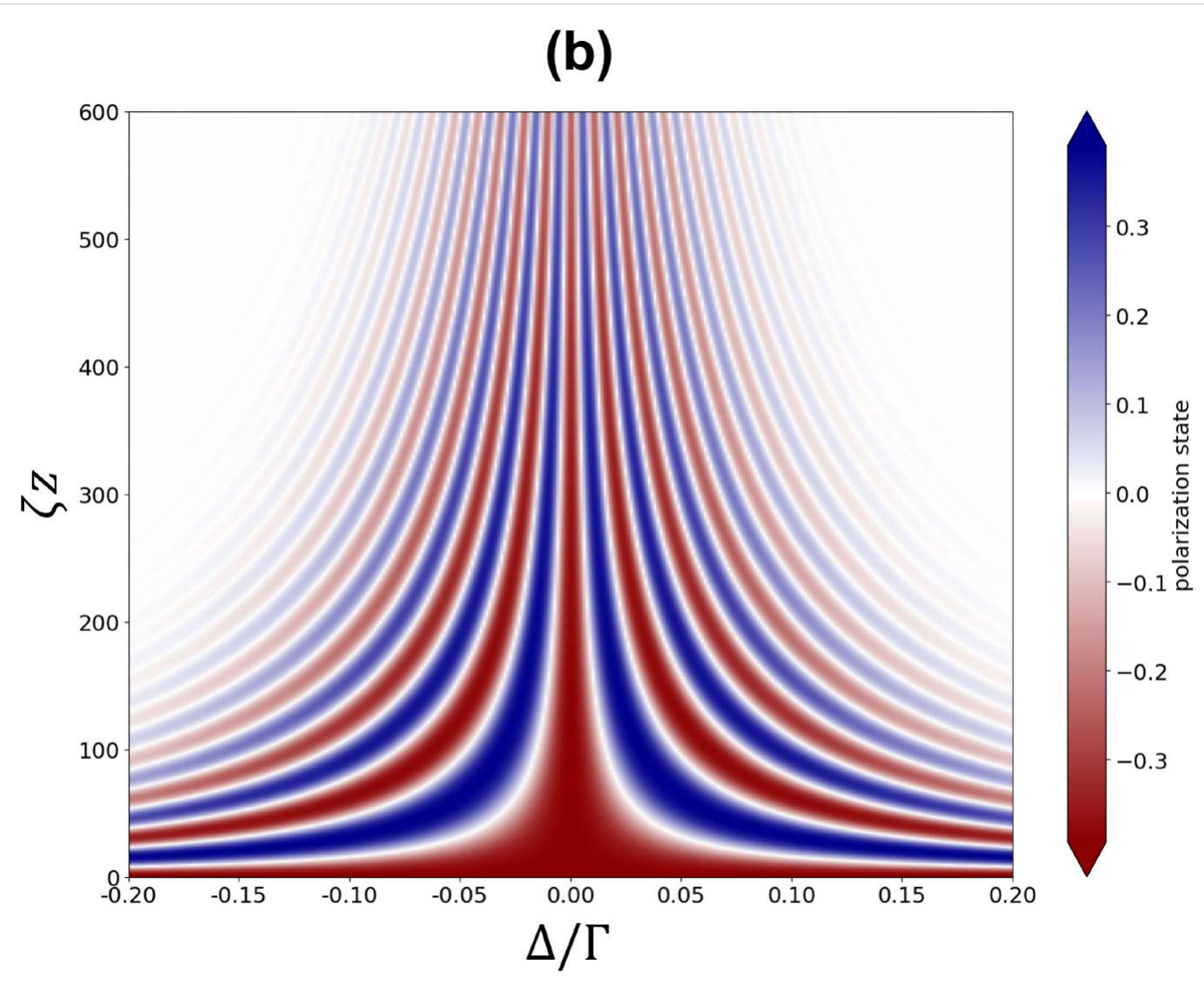}
    \caption{Intensity profile and polarization state evolution (a), and ellipticity distribution (b) of the vector vortex with $|l|=1$ and relative amplitude $\alpha=\pi/8$, such that the initial polarization state at the medium entrance is dominated by the left-handed circular polarization ($|E_L(0)|>|E_R(0)|$). The propagation distance is expressed as the dimensionless parameter $\zeta z$. The relative phase is fixed at $\psi=0$, while the other parameters are $\gamma_d=10^{-3}\Gamma$ and $|\Omega_C|=\Gamma$. (a) The detuning is fixed at $\Delta=0.1\Gamma$; the first, second, and third rows correspond to initial phaseonium states $\theta=\pi/4,\pi/8,3\pi/8$, respectively. Darker rings or lobes correspond to higher intensity regions. The red and blue ellipses denote the left- and right-handed circular polarization states, respectively, while yellow lines indicate linear polarization states. (b) The initial phaseonium state is fixed at $\theta=\pi/4$; the plot shows the average ellipticity corresponding to linear, left- and right-handed circular polarization states, denoted by white, dark red, and dark blue colors, respectively, at different propagation distances $\zeta z$ and detuning values $\Delta/\Gamma$.}
    \label{Polarization varied coherence 1}
\end{figure*}

To quantify the polarization structure of the vector vortex beam across its transverse plane during propagation, we use the Stokes parameters defined in the circular polarization basis \cite{Tarak.OE2022} 
\begin{subequations}
\label{eq.29}
\begin{align}
    S_0 &= |E_R|^2 + |E_L|^2, \label{eq.29a} \\
    S_1 &= 2\mathrm{Re}(E_R^*E_L), \label{eq.29b}\\
    S_2 &=2\mathrm{Im}(E_R^*E_L), \label{eq.29c}\\
    S_3 &=|E_R|^2-|E_L|^2. \label{eq.29d}
\end{align}
\end{subequations}
These parameters allow us to determine the polarization state and texture at each point in the beam. Specifically, the ellipticity is given by $\kappa=\frac{1}{2}\sin^{-1}{(\frac{S_3}{S_0})}$, and the orientation angle is $\xi=\frac{1}{2}\tan^{-1}{(\frac{S_2}{S_1})}$. The orientation angle $\xi$ describes the local rotation of the polarization ellipse, such that after propagating a distance $\zeta z$, the ellipse rotates by $\Delta\xi=\xi(\zeta z)-\xi(0)$. Meanwhile, the ellipticity $\kappa$ classifies the local polarization state: linear polarization for $\kappa=0$, circular polarization for $\kappa=\pm\pi/4$, and elliptical polarization when $0 < |\kappa| <\pi/4$.

We first consider the resonant case with $\Delta=0$, choosing the amplitude tuning angle of the input beam as $\alpha=\pi/8$, so that the initial polarization is dominated by the left-handed circular component. The control field is present with a strength of $|\Omega_C|=\Gamma$, ensuring slow-light propagation of the vector vortex beam. Figure \ref{Polarization Fixed Coherence}(a) corresponds to the situation where the atomic population is set to $\theta=\pi/4$, meaning both ground states $\ket{1}$ and $\ket{2}$ are equally populated before interacting with the vortex components and the control field. At the medium entrance $\zeta z=0$, three distinct polarization textures are observed depending on the relative phase $\psi$; radial ($\psi=0$), spiral $\psi=\pi/2$, and azimuthal $\psi=\pi$, all with a ring-shaped intensity profile corresponding to the topological charge $|l|=1$. As the beam propagates through the medium, the intensity profile gradually evolves from a ring into a two-lobed petal pattern, with the rotation angle of the lobes determined by the relative phase $\psi$. Interestingly, once the petal structure is fully formed and the evolution reaches a quasi-stationary state, the polarization textures convert from the initial left-handed circular polarization (red ellipses) to linear polarization states (yellow lines). This transformation occurs for all values of $\psi$, indicating that the transition from circular to linear polarization is largely independent of the initial relative phase.  

In Figure \ref{Polarization Fixed Coherence}(b), we select $\psi=0$, while the initial atomic population is varied through the tuning angle $\theta$. The first row corresponds to $\theta=\pi/4$, representing equally populated ground states $\ket{1}$ and $\ket{2}$. The second row corresponds to $\theta=\pi/8$, where $\ket{1}$ is initially more populated than $\ket{2}$, and the third row corresponds to $\theta=3\pi/8$, where $\ket{1}$ is initially less populated than $\ket{2}$. As the slow-light vector vortex propagates, the intensity profile gradually transforms from a ring into a petal structure. Since the relative phase $\psi$ is kept constant, no rotation of the lobes is observed for any of the cases. Notably, the stationary polarization states at the end of propagation depend solely on the initial population distribution: for $\theta=\pi/4$, the beam reaches a linear polarization state (yellow lines); for $\theta=\pi/8$, it retains left-handed circular polarization (red ellipses); and for $\theta=3\pi/8$, it evolves into right-handed circular polarization (blue ellipses) for very large propagating distances. These results demonstrate that the final polarization textures of slow-light vector vortices can be directly controlled through the initial superposition population.

Next in Figure \ref{Polarization varied coherence 1}(a), we introduce a small detuning from resonance, setting $\Delta=0.1\Gamma$, while keeping the control field strength $|\Omega_C|$ and other parameters the same as in Fig. \ref{Polarization Fixed Coherence}(b). In this near-resonant regime, the beam no longer experiences perfect EIT, and the dramatic suppression of absorption at $\Delta=0$ is slightly reduced. Nevertheless, the system still exhibits minimal absorption and a positive slope of dispersion, as illustrated in the subplot of Fig. \ref{Absorption at z=0}, ensuring effective slow-light propagation of the vector vortex beam. The small detuning introduces oscillatory behavior in the polarization states at intermediate propagation distances, before the stationary state is reached. During this stage, the intensity profile is not yet fully transformed into lobes, as observed at $\zeta z=100$ and $\zeta z=500$. These oscillations arise from the spatially dependent, complementary gain-loss relationship between the right- and left-handed components, as described in Fig. \ref{Beam Intensity and Susceptibility Evolution}(b). As propagation continues, the gain-loss patterns evolve dynamically, rotating in opposite directions for the two components and varying in size. This dynamic evolution leads to continuous transitions of the local polarization state in the transverse plane, cycling between left-handed circular, linear, and right-handed circular polarizations at different azimuthal angles. Notably, a complete conversion from left- to right-handed circular polarization can occur at intermediate propagation distances while the ring-shaped intensity profile remains largely preserved, as seen for $\theta=\pi/4$ at $\zeta z=100$, and at very large propagation distances, $\zeta z=2000$, for $\theta=3\pi/8$, when the intensity has transformed into lobes. This behavior demonstrates that slow-light vector vortices maintain robust polarization dynamics even under small detuning from resonance.

\begin{figure*}[!t]
    \centering
    \includegraphics[width=0.48\linewidth]{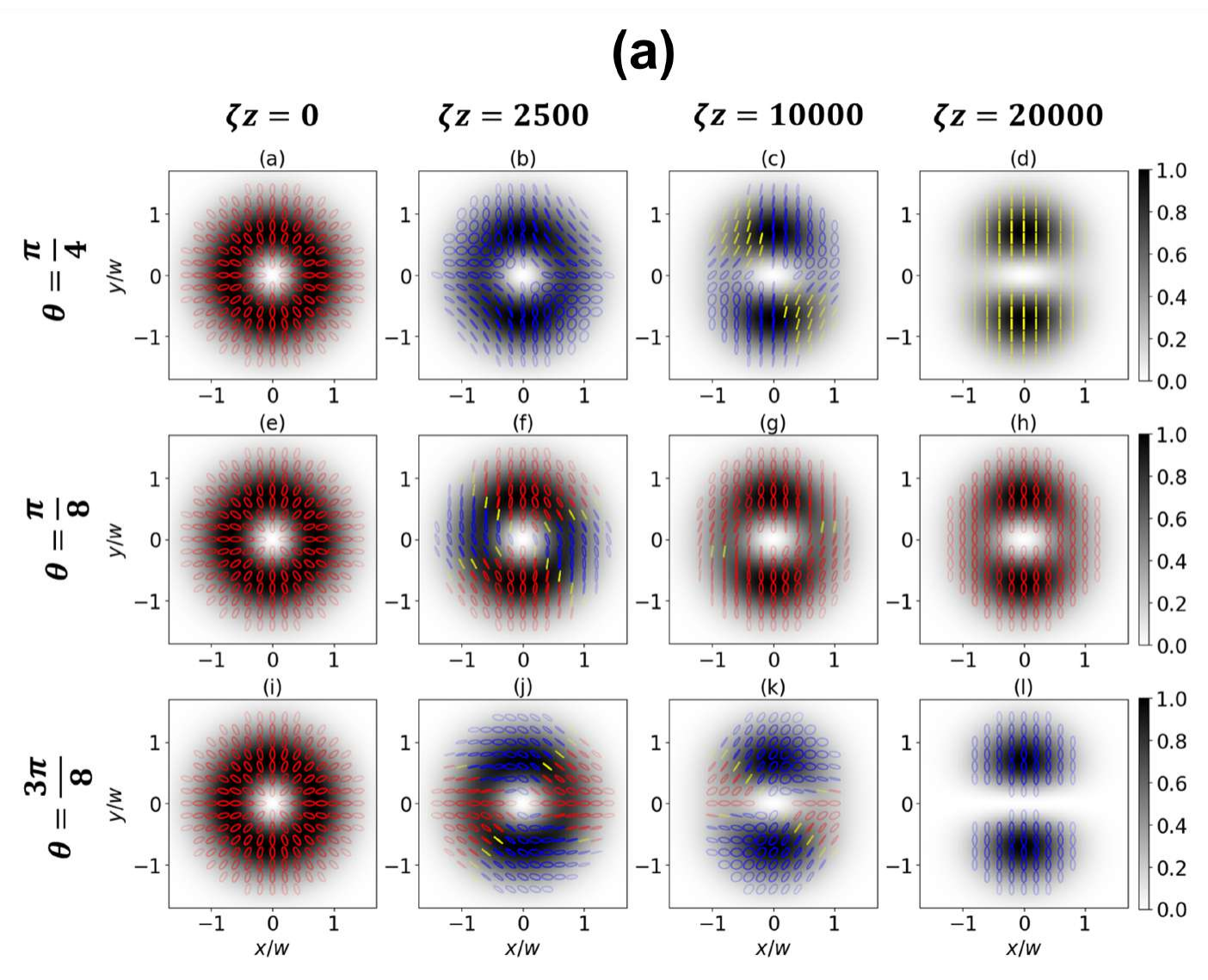}
    \hfill
    \includegraphics[width=0.48\linewidth]{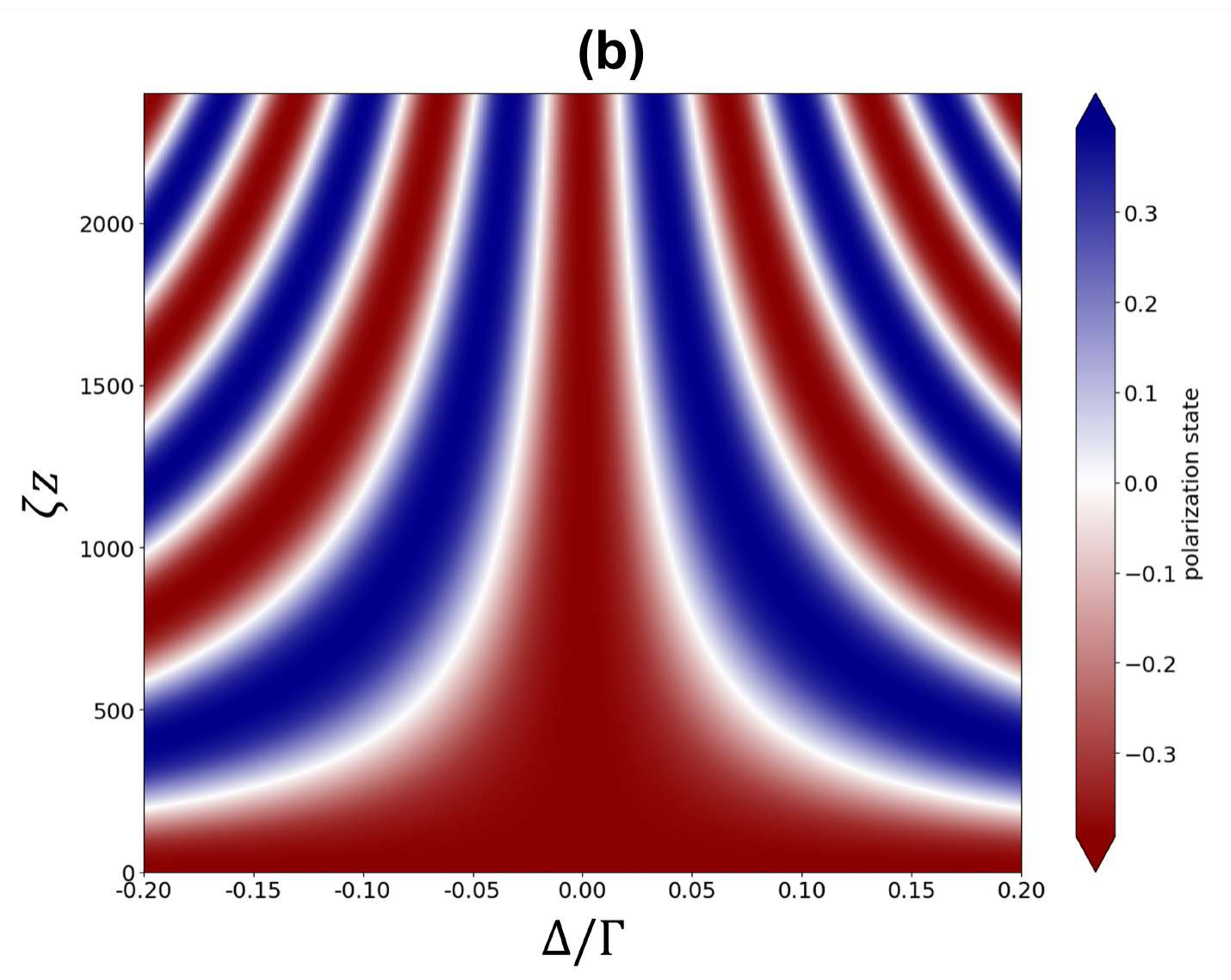}
    \caption{ Intensity profile and polarization state evolution (a), and ellipticity distribution (b) of the vector vortex with $|l|=1$ and relative amplitude $\alpha=\pi/8$,  such that the initial polarization state at the medium entrance is dominated by the left
handed circular polarization ($|E_L(0)|>|E_R(0)|$).  The propagation distance is expressed as the dimensionless parameter $\zeta z$. The relative phase is fixed at $\psi=0$ while the other parameters are $\gamma_d=10^{-3}\Gamma$ and $|\Omega_C|=5\Gamma$. (a) The detuning is fixed at $\Delta=0.1\Gamma$; the first, second, and third rows correspond to initial phaseonium states  $\theta=\pi/4,\pi/8,3\pi/8$, respectively. Darker
rings or lobes correspond to higher intensity regions. The red and blue ellipses denote the left- and right-handed circular
polarization states, respectively, while yellow lines indicate linear polarization states. (b)  The initial phaseonium state is fixed at $\theta=\pi/4$;  the plot shows the average ellipticity corresponding to linear, left- and right-handed circular polarization states,
denoted by white, dark red, and dark blue colors, respectively, at different propagation distances  $\zeta z$ and detuning values  $\Delta/\Gamma$.}
    \label{Polarization varied coherence 2}
\end{figure*}

To illustrate the continuous and periodic transitions of the polarization states before the system reaches a stationary regime, we plot the average ellipticity across the beam cross section for regions with non-zero intensity, at different normalized propagation distances $\zeta z$ and detuning values $\Delta/\Gamma$. The results are shown in Fig. \ref{Polarization varied coherence 1}(b) for $\theta=\pi/4$, where left- and right-handed circular polarizations are indicated in red and blue, respectively, and linear polarization is shown in white to provide clear contrast among the three distinct polarization states. From $\zeta z = 0$ to $\zeta z = 600$, the polarization periodically transitions from left-handed circular to linear and then to right-handed circular for non-zero detuning $\Delta/\Gamma \neq 0$. At longer propagation distances, the system reaches a stationary regime in which the polarization remains linear due to the chosen atomic population parameter $\theta=\pi/4$. The colormap also shows that for larger detuning from resonance, the distance required to reach the stationary state decreases, reflecting stronger medium losses that transform the ring-shaped intensity profile into lobes more quickly, with symmetric behavior for positive and negative detunings.

In Figure \ref{Polarization varied coherence 2}(a), we introduce a stronger control field with $|\Omega_C| = 5\Gamma$, while all other parameters are the same as in Fig. \ref{Polarization varied coherence 1}(a). The stronger control field significantly reduces the total absorption in the system, which slows down the evolution of both the beam’s intensity profile and its polarization states. As a result, the propagation distance required to reach the stationary state is increased by roughly an order of magnitude compared to the small-detuning, weaker control field case shown in Fig. \ref{Polarization varied coherence 1}(a). Figure \ref{Polarization varied coherence 2}(b) presents the average ellipticity across the beam cross section at different normalized propagation distances $\zeta z$ and detuning values $\Delta/\Gamma$, for the atomic parameter $\theta=\pi/4$. For non-zero detuning, the beam remains in nearly constant circular polarization states over a longer propagation distance before transitioning periodically into linear and opposite circular polarization states. This behavior contrasts with the faster oscillations observed for the weaker control field in Fig. \ref{Polarization varied coherence 1}(b). Such a scenario may be advantageous experimentally, as the medium length can be chosen so that polarization flipping occurs at an intermediate propagation distance, where the ring-shaped intensity profile is largely preserved, thereby maintaining the information encoded in the beam’s intensity.

\section{Concluding remark}
In conclusion, we have investigated the propagation of slow-light optical vector vortices in a four-level tripod atomic system. By preparing the medium in a coherent superposition of two ground states and coupling the remaining ground state to the excited state with a strong control field, we demonstrated the emergence of phase-dependent transparency (or gain–loss pairs) with $2|l|$-fold degeneracy and tunable spin–orbit coupling, manifested as polarization evolution during propagation. Analytical solutions of the linear Maxwell–Bloch equations under the paraxial and slowly varying envelope approximations show that the strong control field enables slow-light propagation of the vector vortex with reduced absorption. We further analyzed the role of the atomic population, which determines the final polarization state once the beam reaches a stationary regime. In this regime, the initially ring-shaped intensity profile transforms into a petal-like structure with $2|l|$ lobes, reflecting the phase-dependent transparency (or gain–loss pairs), largely independent of the initial vector vortex parameters. At intermediate propagation distances, while the ring-shaped intensity is still partially preserved, small detunings induce periodic oscillations of the polarization state. The rate of these oscillations can be controlled by the strength of the control field, enabling complete polarization flipping at intermediate distances without destroying the ring-shaped intensity. These results demonstrate the feasibility of controlling both the orbital and spin angular momentum properties of slow-light vector vortices in atomic media, offering potential applications in high-dimensional quantum communication and information processing.

\begin{acknowledgments}
This project has received funding from the Research Council of Lithuania (LMTLT), agreement No. S-ITP-24-6. D.P.P. gratefully acknowledges Dr. Viačeslav Kudriašov for the technical help on simulation of vortices, and gratefully acknowledges the support of the Erasmus+: Erasmus Mundus programme of the European Union under
Convention $\mathrm{n^o}$ 101128124 — EUROPHOTONICS — ERASMUS-EDU-2023-PEX-EMJMMOB. 
\end{acknowledgments}

\bibliography{apssamp.bib}

\end{document}